\documentclass[aps,prd,preprint,longbibliography,onecolumn,showpacs,amsmath,amssymb,superscriptaddress,nofootinbib,floatfix,preprintnumbers,secnumarabic]{revtex4-2}

\usepackage{dcolumn}
\usepackage{bm}
\usepackage[T1]{fontenc}
\usepackage{graphicx}
\usepackage{amssymb,amsmath}
\usepackage{booktabs}
\usepackage{multirow}
\usepackage[dvipsnames]{xcolor}
\def\linkcolor{cyan!70!black}

\usepackage[
colorlinks=true
,urlcolor=\linkcolor
,anchorcolor=\linkcolor
,citecolor=\linkcolor
,filecolor=\linkcolor
,linkcolor=\linkcolor
,menucolor=\linkcolor
,linktocpage=true
,pdfproducer=medialab
,pdfa=true
]{hyperref}
\usepackage{tikz}
\usepackage{mathtools}
\usepackage[capitalise]{cleveref}
\usepackage{cancel}
\usepackage{feynmp}
\usepackage{soul}
\usepackage{comment}

\usepackage[normalem]{ulem}

\usepackage{slashed}
\usepackage{epsfig,psfrag,rotating,soul}
\usepackage{rotfloat}
\usepackage[font={small},justification=Justified, singlelinecheck=false,format=plain]{caption}
\usepackage{colortbl}
\usepackage{soul}
\usepackage{xtab}

\definecolor{lime}{HTML}{A6CE39}
\DeclareRobustCommand{\orcidicon}{%
    \begin{tikzpicture}
    \draw[lime, fill=lime] (0,0) 
    circle [radius=0.16] 
    node[white] {{\fontfamily{qag}\selectfont \tiny ID}};
    \draw[white, fill=white] (-0.0625,0.095) 
    circle [radius=0.007];
    \end{tikzpicture}
    \hspace{-2mm}
}
\newcommand{\orcid}[1]{\href{https://orcid.org/#1}{\orcidicon}}

\newcommand{\be}{\begin{equation}}
\newcommand{\ee}{\end{equation}}

\newcommand{\beq}{\begin{equation}} 
\newcommand{\eeq}{\end{equation}} 
\newcommand{\ba}{\begin{array}}  
\newcommand{\ea}{\end{array}} 
\newcommand{\bea}{\begin{eqnarray}}  
\newcommand{\eea}{\end{eqnarray} }  
\newcommand{\bal}{\begin{align}}
\newcommand{\eal}{\end{align}}   
\newcommand{\bi}{\begin{itemize}}  
\newcommand{\ei}{\end{itemize}}  
\newcommand{\ben}{\begin{enumerate}}
\newcommand{\een}{\end{enumerate}}  
\newcommand{\bc}{\begin{center}}
\newcommand{\ec}{\end{center}} 
\newcommand{\bt}{\begin{table}}
\newcommand{\et}{\end{table}}  
\newcommand{\btb}{\begin{tabular}}
\newcommand{\etb}{\end{tabular}}

\allowdisplaybreaks

\let\OLDthebibliography\thebibliography
\renewcommand\thebibliography[1]{
  \OLDthebibliography{#1}
  \setlength{\parskip}{0pt}
  \setlength{\itemsep}{0pt plus 0.3ex}
}

\usepackage{fontawesome5}
\makeatletter
\newcommand{\github}[1]{%
   \href{#1}{\faGithubSquare}%
}
\makeatother

\begin{document}

\begin{titlepage}

\begin{flushright}
    IFT-UAM/CSIC-24-172
\end{flushright}


\title{
Probing a diffuse flux of axion-like particles from galactic supernovae with neutrino water Cherenkov detectors
}


\author{David Alonso-Gonz\'alez\orcid{0000-0002-7572-9184}}
\email{david.alonsogonzalez@uam.es}
\affiliation{Instituto de F\' \i sica Te\'orica, IFT-UAM/CSIC, 28049 Madrid, Spain}
\affiliation{Departamento de F\' \i sica Te\'orica, Universidad Aut\'onoma de Madrid, 28049 Madrid, Spain}

\author{David Cerde\~no\orcid{0000-0002-7649-1956}}
\email{davidg.cerdeno@gmail.com}
\affiliation{Instituto de F\' \i sica Te\'orica, IFT-UAM/CSIC, 28049 Madrid, Spain}

\author{Marina~Cerme\~no\orcid{0000-0001-6881-7285}}
\email{marina.cermeno@ift.csic.es}
\affiliation{Instituto de F\' \i sica Te\'orica, IFT-UAM/CSIC, 28049 Madrid, Spain}

\author{Andres D. Perez\orcid{0000-0002-9391-6047}}\email{andresd.perez@uam.es}
\affiliation{Instituto de F\' \i sica Te\'orica, IFT-UAM/CSIC, 28049 Madrid, Spain}
\affiliation{Departamento de F\' \i sica Te\'orica, Universidad Aut\'onoma de Madrid, 28049 Madrid, Spain}

\date{\today}
\begin{abstract}
In this article, we claim that axion-like particles (ALPs) with MeV masses can be produced with semi-relativistic velocities in core-collapse supernovae (SNe), generating a diffuse galactic flux. We show that these ALPs can be detected in neutrino water Cherenkov detectors via $a \, p \rightarrow p \, \gamma$ interactions. Using Super-Kamiokande data, we derive new constraints on the ALP parameter space, excluding a region spanning one order of magnitude in the ALP-proton coupling above cooling bounds for ALP masses in the range of $1-70$~MeV and ALP-proton couplings between $\sim 2\times10^{-5}-2\times10^{-4}$. We show that the future Hyper-Kamiokande will be able to probe couplings as small as $\sim 10^{-5}$, considerably constraining the allowed region abov
e SN 1987A cooling bounds.
\end{abstract}

\preprint{IFT-UAM/CSIC-24-172}


\maketitle
\thispagestyle{empty}

\end{titlepage}

\section{Introduction}\label{sec:intro}

Axions and axion-like particles (ALPs) are pseudo-scalar fields arising in well-motivated Standard Model (SM) extensions~\cite{DiLuzio:2020wdo}. It is well known that axions emerge as pseudo-Nambu Goldstone bosons in the Peccei-Quinn mechanism, which solves the strong CP-problem~\cite{Peccei:1977hh1, Peccei:1977hh2, Weinberg:1977ma, Wilczek:1977pj, Arvanitaki:2009fg}. Similarly, ALPs are pseudo-Nambu Goldstone bosons predicted in several beyond the SM theories~\cite{Svrcek:2006yi, Gelmini:1980re, Davidson:1981zd, Wilczek:1982rv, Cicoli:2012sz, Halverson:2019kna}, where new global symmetries are introduced and subsequently broken at a high energy scale, $f_a$. Although they do not solve the strong CP-problem, ALPs share many of the phenomenological properties of axions. Their interactions with SM particles are suppressed by the high energy scale, leading to very weak couplings. This makes ALPs viable as dark matter (DM) candidates and as mediators of new interactions~\cite{Preskill:1982cy, Abbott:1982af, Dine:1982ah}. In particular, MeV-scale ALPs have recently received significant attention, as they can act as mediators between DM and SM particles, reproducing the total expected DM relic abundance via thermal freeze-out~\cite{Boehm:2014hva,Hochberg:2018rjs}.

Core-collapse supernovae (SNe) serve as powerful natural laboratories for testing ALPs, as these particles can be copiously produced in their dense and hot interiors if they couple to the SM constituents of SN matter. This phenomenon has been extensively studied in the literature to impose stringent constraints on the parameter space of various ALP models~\cite{RAFFELT19901, Raffelt:1993ix, Raffelt:1996wa, Fischer:2016cyd, Carenza:2019pxu, Carenza:2020cis, Fischer:2021jfm, Lella:2022uwi, Lucente:2022vuo,  Lella:2023bfb, Carenza:2023lci,  Chakraborty:2024tyx, Lella:2024dmx}. Notably, the ALP-nucleon coupling has been constrained using data from SN 1987A. Nevertheless, ALP masses above $1$ MeV remain largely unconstrained, apart from cooling bounds~\cite{Lella:2023bfb, Carenza:2023lci}, which cannot probe couplings above $g_{aN} \sim 10^{-6}$.

In this article, we focus on this parameter region, where ALPs are produced in the proto-neutron star (proto-NS) interior through nucleon-nucleon Bremmstrahlung, $N\, N\, \rightarrow N\, N\, a$, and pion-ALP conversions, $N\, \pi \rightarrow N\, a$, but the SN environment becomes optically thick to ALPs, preventing them from free-streaming. Even if a significant fraction of these particles leaves the proto-NS core, they fail to release sufficient energy to impact the SN cooling mechanism. ALPs in this regime would escape with semi-relativistic velocities, reaching Earth over a time period of $10^3$~years.

We argue that these MeV ALPs from galactic SNe form a diffuse galactic flux. We demonstrate that they can produce detectable signals in neutrino water Cherenkov detectors through interactions with free protons via $a \, p \rightarrow p \, \gamma$. This allows us to probe previously unexplored regions of the ALP parameter space and derive constraints using Super-Kamiokande (SK) data~\cite{Super-Kamiokande:2021jaq}. Focusing on a model inspired by the Kim-Shifman-Vainshtein-Zakharov (KSVZ) axion model~\cite{Kim:1979if, Shifman:1979if}, we exclude a region spanning one order of magnitude in the ALP-proton coupling above the cooling bounds for ALP masses in the range of $1-70$~MeV. Additionally, we present projected bounds for Hyper-Kamiokande (HK)~\cite{Hyper-Kamiokande:2018ofw}, which would improve upon the SK constraints by up to a factor of $\sim 2.5$, considerably constraining the region above the upper cooling limit.

This article is organized as follows. In Section~\ref{sec:ALP-prod} we derive the ALP production spectrum in core-collapse SN and the spectral fluence of ALPs arriving at Earth. In Section~\ref{sec:diffuse} we advocate for the existence of a diffuse galactic SN flux of ALPs for $m_a \gtrsim 1$ MeV, discuss its behaviour with the ALP mass and coupling to protons and show examples for some benchmark points. In Section~\ref{sec:cherenkov}, we analyse the expected photon spectrum in Cherenkov neutrino detectors from interactions of the diffuse flux of ALPs on free protons. We derive the  excluded ALP parameter space by combining the first four phases of SK, and present the projected bounds for HK. Finally, our conclusions are presented in Section~\ref{sec:conclusions}.

\section{ALPs from SN}\label{sec:ALP-prod}

Core-collapse SNe reach temperatures up to $T \sim 30$ MeV, and baryonic densities several times higher than the nuclear saturation density, $\rho_0 = 3\times 10^{14} \, \rm g/cm^3$, within the first seconds after the explosion \cite{fischer2012, Fischer:2021jfm}. These extreme conditions provide an ideal environment for ALP production.

The ALP-nucleon interaction can be described by the following Lagrangian~\cite{PhysRevD.40.652,DiLuzio:2020wdo,Chang:1993gm, Lella:2024dmx}
\begin{align}
    \mathcal{L}_{int} = g_a \, \frac{\partial_{\mu} a}{2 \, m_N} \, \bigg[ 
    & C_{ap} \, \bar{p} \, \gamma^{\mu} \, \gamma_5 \, p 
    + C_{an} \, \bar{n} \, \gamma^{\mu} \, \gamma_5 \, n \notag \\
    & + \frac{C_{a \pi N}}{f_{\pi}} \, \left( i \, \pi^+ \, \bar{p} \, \gamma^{\mu} \, n - i \, \pi^- \, \bar{n} \, \gamma^{\mu} \, p \right) \notag \\
    & + C_{a N \Delta} \, \left( \bar{p} \, \Delta^+_{\mu} + \bar{\Delta^+_{\mu}} \, p 
    + \bar{n} \, \Delta^0_{\mu} + \bar{\Delta^0_{\mu}} \, n \right) 
    \bigg],
    \label{eq:Laxion}
\end{align}
where $p$ is the proton, $n$ the neutron, $\pi$ the pion, and $\Delta$ the baryon that represents the $\Delta$-resonance. We introduce a dimensionless coupling constant related to the high energy ALP scale as $g_a = m_N / f_a$, where $m_N$ is the nucleon mass, and $C_{aN}$, with $N=p, n$, are model-dependent $\mathcal{O}(1)$ axion-nucleon coupling constants. The pion decay constant is $f_{\pi}=92.4$~MeV, and the axion-pion-nucleon and axion-nucleon-$\Delta$ baryon coupling constants depend on the previous as $C_{a \pi N} = (C_{ap} - C_{an}) / \sqrt{2} g_A$~\cite{Choi:2021ign},  $C_{a N \Delta} = - \sqrt{3}/2 (C_{ap} - C_{an})$~\cite{Ho:2022oaw}, with $g_A \simeq 1.28$~\cite{PDG:2022} the axial coupling.

The two terms on the first line of Eq.~\eqref{eq:Laxion} represent the ALP-nucleon vertex, which gives rise to the dominant process responsible for ALP emission from hot nuclear matter, nucleon-nucleon bremsstrahlung, $N N \to N N a$~\cite{RAFFELT19901, PhysRevD.38.2338,Raffelt:1993ix, Giannotti:2005tn, Carenza:2019pxu}. We define the ALP-proton and ALP-neutron coupling as $g_{aN} = g_a \, C_{aN}$, for convenience. The second and third lines contribute to the Compton-like scattering, $\pi\, N \rightarrow N\, a$, also referred to as pion-ALP conversion process, relevant for ALP production in SN for $m_a \gtrsim 135$~MeV. In particular, the second line  represents a four-particle interaction vertex, which corresponds to a contact interaction between the ALP, a pion, and two nucleons~\cite{PhysRevLett.60.1793,PhysRevLett.60.1797,PhysRevD.39.1020,PhysRevD.42.3297,Carenza:2019pxu,Carenza:2020cis,Fischer:2021jfm,Choi:2021ign}. The third line describes the ALP-$\Delta$-resonance vertex that is involved in pion-ALP conversion through an intermediate virtual state, either a nucleon or a $\Delta$-resonance~\cite{Ho:2022oaw}.

Note that the terms on the first line of Eq.~\eqref{eq:Laxion} also lead to the photo-production channel $\gamma \, p \rightarrow a \, p$. However, as it can be found in Appendix~\ref{app:production}, its contribution to the ALP production rate is at least three orders of magnitude smaller than nucleon-nucleon Bremmstrahlung and pion-axion conversion.

The general formula for the ALP production spectrum (number density of ALPs produced per unit time and energy in the proto-NS) is~\cite{RAFFELT19901}
\begin{equation}
    \frac{d^2n_a}{dE_a dt}  =  \prod_i \int \frac{g_id^3 \vec{p}_i}{(2 \pi)^3 2 E_i}   \prod_{j \neq a} \int \frac{d^3 \vec{p'}_j}{(2 \pi)^3 2 E'_j}  \, (2 \pi )^4 \delta^4 \Bigl( \sum_k p_k - \sum_{l \neq a} p'_l-p_a \Bigr) \frac{|{\vec{p}_a}|}{4 \pi^2} |{\mathcal{\overline{M}}|}^2 \mathcal{F}(E_i, E'_j),    
    \label{eq:prod_spectrum}
\end{equation}
where $p_i=(E_i,\vec{p}_i)$ are the four-momenta of the incoming particles, with $E_i$ and $\vec{p}_i$ representing their energies and three-momenta, respectively, and $g_i$ are their degrees of freedom. Similarly, $p'_j=(E'_j,\vec{p'}_j)$ denote the four-momenta of the outgoing particles, except for the ALP. The four-momentum of the produced ALP is $p_a=(E_a,\vec{p}_a)$, where $E_a$ and $\vec{p}_a$ correspond to its energy and three-momentum. The term $|{\mathcal{\overline{M}}|}^2$ is the squared spin-averaged matrix element for the processes producing ALPs. The function $\mathcal{F}(E_i, E'_j)$ describes the energy distribution of the particles involved in the interactions and is given by $\mathcal{F}(E_i, E'_j)=f_i (E_i)[1 - f_j (E'_j) ]$. Here, $f_i (E_i)$ and $f_j (E'_j)$ are the distribution functions for the incoming and outgoing particles, respectively, which follow a Fermi-Dirac distribution for nucleons and a Bose-Einstein distribution for pions.

As it has been already mentioned, ALPs can be produced in the SN core by nucleon-nucleon bremsstrahlung and by pion conversions, specifically $\pi^- \,p \rightarrow n\, a$ and $\pi^0 \, n \rightarrow n\, a$, since the amount of $\pi^+$ is negligible in the proto-NS interior~\cite{Fore:2019wib, Fischer:2021jfm,  Choi:2021ign}. The expressions for the ALP production spectrum for these specific processes can be found in Ref.~\cite{Carenza:2023lci}.

If ALPs are weakly-coupled to matter, $g_{aN} \lesssim 10^{-8}$, they leave the star unimpeded after being produced. This is referred to as the free-streaming regime. This regime is already tightly constrained by cooling bounds, see Refs.~\cite{Lella:2022uwi, Lella:2023bfb, Carenza:2023lci}. 
Here, we focus on the trapping regime, where absorption effects via $N\, N\, a \rightarrow N\, N$ and $N\, a \rightarrow N\, \pi$ are important, and ALPs cannot free-stream out of the star (in Appendix~\ref{app:production}, we show that the impact of $a\, p \rightarrow \gamma\, p$ on the absorption process is negligible). In this regime, the integrated ALP spectrum over a spherically symmetric SN profile can be computed as~\cite{Lella:2023bfb}
\begin{equation}
    \frac{d^2 N_a}{d E_a d t}=\int_0^{\infty} 4 \pi r^2 d r\left\langle e^{-\tau\left(E_a^*, r\right)}\right\rangle \frac{d^2 n_a}{d E_a d t}\left(E_a, r\right),
    \label{eq:spectrum}
\end{equation}
where $r$ is the radial position with respect to the center of the SN core, and $\tau\left(E_a^*, r\right)$ is the optical depth at a given ALP energy and position.

The exponential term 
\begin{equation}
\left\langle e^{-\tau\left(E_a^*, r\right)}\right\rangle=\frac{1}{2} \int_{-1}^{+1} d \mu \, e^{-\int_0^{\infty} d s \Gamma_a\left(E_a^*, \sqrt{r^2+s^2+2 r s \mu}\right)}
\label{eq:expabs}
\end{equation}
encodes the absorption effects over the ALP emission and it is obtained by averaging over the cosine of the emission angle $\mu=\rm cos\;\theta$, with $\theta$ the angle between the ALP propagation direction and the radial vector $\vec{r}$ from the origin to the point of ALP production.
The absorption rate $\Gamma_a$ can be obtained as
\begin{equation}
\Gamma_a\left(E_a, r\right)=\lambda_a^{-1}\left(E_a, r\right)\left[1-e^{-\frac{E_a}{T(r)}}\right],
\label{eq:abs}
\end{equation}
where $T(r)$ is the SN temperature profile and $\lambda_a\left(E_a, r\right)$ is the ALP mean free path in the proto-NS core, see Appendix A of Ref.~\cite{ Carenza:2023lci} for details of this computation.

The exponential term given by Eq.~\eqref{eq:expabs} is evaluated at the ALP energy 
\begin{equation}
E_a^*=E_a \, \frac{\alpha(r)}{\alpha\left(\sqrt{r^2+s^2+2rs\mu}\right)},
\label{eq:eared}
\end{equation}
which takes into account the gravitational redshift from the point of ALP production to the point of absorption through the lapse function $\alpha(r) \leq 1$, that encodes effects due to the proto-NS gravitational potential, $\Phi(r)$, evaluated locally in the proto-NS interior.
In a similar way to the derivation of the gravitational redshift in Ref.~\cite{DeRocco:2019jti}, we can write the lapse function as
\begin{equation}
\alpha(r)=\sqrt{1-2\Delta\Phi(r)},
\label{eq:lapse}    
\end{equation}
where 
\begin{equation}
\Delta \Phi (r)=G \int_{r}^{\infty} \frac{m_{\mathrm{enc}}(r')}{r'^2} d r'
\label{eq:pot} 
\end{equation}
is the change in potential between $r$ and $\infty$ and $m_{\rm enc}(r')$ is the mass enclosed in $r'$.

The spectral fluence of ALPs resulting from an isotropic production spectrum at a detector located far from the SN can be expressed as  
\begin{equation}
\frac{dN_a}{dE_a^{* \infty}} = \int_{t_{\rm min}}^{t_{\rm max}} dt \int_0^{\infty} \alpha(r)^{-1} 4 \pi r^2 dr \, \left\langle e^{-\tau\left(E_a^*, r\right)} \right\rangle \frac{d^2 n_a}{dE_a dt} \left( r, t, \alpha(r)^{-1} E_a^{* \infty} \right),
\label{eq:fluence}    
\end{equation}  
where $E_a^{* \infty}=\alpha(r) E_a$ is the observed energy at infinity, redshifted relative to the local energy $E_a$. The integration over time spans from $t_{\rm min}$ to $t_{\rm max}$, which correspond to the times, relative to the moment of core collapse, during which the bulk of ALP production occurs.

\section{Diffuse galactic flux of ALPs}
\label{sec:diffuse}

ALPs with masses above $m_a \sim 1$ MeV would be produced in core-collapse SN with semi-relativistic velocities, constituting a diffuse galactic flux. The reason is that they are emitted traveling with an $\mathcal{O}(1)$ spread in velocities $\Delta v$ and, therefore, similarly to the argument used in Ref.~\cite{DeRocco:2019jti} for dark fermions, this results in a difference in arrival time of the high-velocity and low-velocity ALPs. This difference depends on the distance $d$ to the SN and can be estimated as $\Delta t \simeq d / \bar{v} \, \delta v$, with $\bar{v}$ the average ALP velocity and $\delta v= \Delta v / \bar{v}$.

If a SN is located at the Galactic Center (GC), $d=8.7$ kpc, ALPs with masses $m_a \sim 1- 100$~MeV arrive at Earth in a time window of $\Delta t \sim 5 \times 10^2- 5\times 10^3$ years. Since the SN rate in our Galaxy is $\sim 2$ per century~\cite{Beacom:2010kk,Adams:2013ana}, we expect that the flux of $\sim 10-100$ SN overlap at Earth, resulting in a near-constant galactic flux at any time. This is in contrast with SN neutrinos, which are produced with the speed of light $c$, causing that the entire flux arrives within the $\sim 10$ second window of production, avoiding a constant neutrino flux from galactic origin. Note that, in our case, contrary to the diffuse SN neutrino background (DSNB), which is extragalactic, the diffuse galactic ALP flux is anisotropic, peaked towards the GC, where the SN rate is larger.

The ALP flux from galactic SNe that reaches Earth ($  E_a^{\mathrm{Earth}}\equiv E_a^{* \infty}$) can be written as
\begin{equation}
\frac{d\Phi_a}{dE_a^{\mathrm{Earth}}} = 	\frac{dN_a}{dE_a^{\mathrm{Earth}}} \int_{-\infty}^{\infty} \int_0^{2\pi} \int_0^{\infty} \frac{dn_{SN}}{dt} \, \frac{r}{4 \pi \, (\vec{r} - \vec{R}_E)^2} \, dr \, d\theta \, dz \,, 
  \label{eq:alp_flux}
\end{equation}
where
\begin{equation}
    \frac{dn_{SN}}{dt} = A \, \mathrm{e}^{-\frac{r}{R_d}} \, \mathrm{e}^{-\frac{|z|}{H}},
\label{eq:SNrate}
\end{equation}
is the galactic SN rate~\cite{Adams:2013ana}, $|\vec{R_E}|=8.7$ kpc is the Earth distance with respect to the GC and $z_E=24$ pc its position above the mid-plane of the disk. For Type II SN, $R_d=2.6$~kpc, $H=300$ pc~\cite{McMillan:2016jtx} and the normalization factor is $A=6.40\times10^{-4}$ kpc$^{-3}$ yr$^{-1}$, since we expect a galactic SN rate of $1.63 \pm 0.46$ events per century~\cite{Rozwadowska:2020nab}. This results in
\begin{equation}
   \frac{d\Phi_a}{dE_a^{\mathrm{Earth}}} = 1.3 \times 10^{-55} \; \mathrm{cm^{-2} \; s^{-1}} \, \frac{dN_a}{dE_a^{\mathrm{Earth}}},  \label{eq:alp_flux2} 
\end{equation}
with a $\sim 30 \%$ uncertainty given by current errors in the SN rate and distribution.

It is important to remark that, for $m_a < 1$ MeV and $\Delta t < 5 \times 10^2$ years, we expect a flux of less than $10$ SN overlapping at Earth. In that regime, the assumption of a diffuse flux of ALPs begins to break down, as the arrival times of ALPs from individual SNe would not overlap sufficiently, resulting in a time-dependent galactic flux rather than a steady one. Therefore, we restrict ourselves to $m_a \geq 1$ MeV.

Note also that the diffuse galactic flux is only possible for long-lived ALPs, with decay length $\lambda_a \geq \mathcal{O}(d_g)$, with $d_g$ the diameter of the galaxy. This is possible in some general high-energy ALP theories, where low-energy ALP-photon terms may appear but the total coupling is suppressed~\cite{GrillidiCortona:2015jxo,Craig_2018,Lella:2024dmx}.

For concreteness, we fix $C_{ap}=-0.47$ and $C_{an}=0$, i.e., $g_{an}=0$, in analogy with the KSVZ axion model~\cite{Kim:1979if, Shifman:1979if}, which has already being studied in SN contexts, see for example Refs.~\cite{Lella:2022uwi, Lella:2023bfb, Carenza:2023lci}.

To describe the SN environment, we consider a model with a $18 M_\odot$ progenitor simulated in spherical symmetry with the AGILE-BOLTZTRAN code \cite{agile1,agile2}. In particular, we use the temperature, density, electron and muon fraction profiles of Ref.~\cite{Fischer:2021jfm} at 1 second after bounce. To get the ALP spectrum, we integrate over a time window of $1.5$ seconds, i.e, $t_{\rm min}=0.5$~s and $t_{\rm max}=2$~s in Eq.~\eqref{eq:fluence}, since these profiles are not expected to significantly change in that time frame (see, for example, Fig.~7 in Ref.~\cite{fischer2012}). This will provide conservative bounds once we compare the spectrum of expected events from ALPs in neutrino detectors with the background from SM processes. 

The SN profiles considered in this work assume spherical symmetry (1D). However, as it was shown using 3D simulations~\cite{Tamborra:2014aua,Vartanyan:2019ssu}, the evolution of a SN event can be anisotropic, leading to variations of up to 20$\%$ in the emission of neutrinos, depending on the angle of emission. Although further dedicated studies are required to fully account for this effect in the ALP case, a similar uncertainty range in its flux can be expected. Since the diffuse galactic flux of ALPs is due to the overlap of several SN contributions, the resulting flux would be the average on the emission angle. This average was shown to be similar for 1D and 3D cases in Ref.~\cite{Janka:2025tvf}.

For simplicity, we have assumed that the profiles of all galactic SNe are equivalent to that of a core-collapse SN from an $18M_\odot$ progenitor. The majority of core-collapse SN are expected to originate from progenitor stars with masses below $18M_\odot$ (see Fig. 14 of Ref.~\cite{Sukhbold:2015wba}). One second after bounce, the maximum temperature and density of core-collapse SN from these progenitors are lower than those expected for the $18M_\odot$ case. In particular, for an $8M_\odot$ progenitor, these values are predicted to be $20\% $ lower than those considered in our model (see Fig. 1 of Ref.~\cite{Calore:2021hhn}). Although only $15\% $ of the total SN are expected to have a progenitor mass above $18M_\odot$ these can be significantly hotter and denser. To assess the expected variation in the ALP flux, a study considering the progenitor mass distribution and the simulations of all the SN profiles for each mass would be needed, however that analysis is outside the scope of this work.

Finally, in Ref.~\cite{Fischer:2021jfm}, it was shown that for massless ALPs the SN profiles are not affected. For simplicity, we assume that this is also the case for our massive ALPs.

\begin{figure}[!t]
  \centering
  \includegraphics[width=\textwidth]{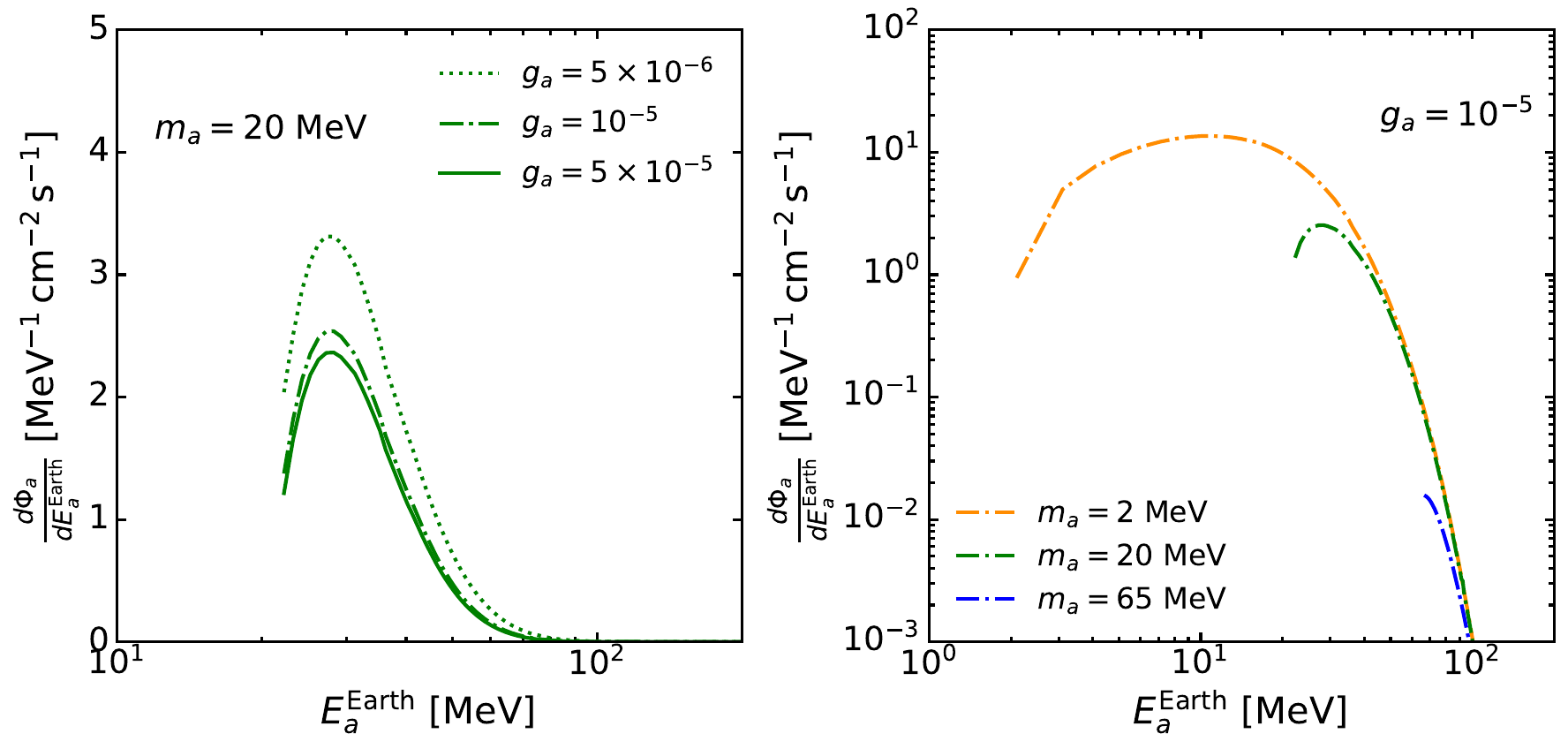}
  
  \caption{ALP flux from galactic SNe that reaches Earth as a function of the ALP energy. In the left plot, for $m_a=20$ MeV and $g_a=5\times 10^{-6}, \, 10^{-5}, \, 5\times 10^{-5}$ in dotted, dashed-dotted and solid lines, respectively. In the right plot, $g_a=10^{-5}$ is fixed and $m_a=2, \, 20, \, 65$ MeV, in orange, green and blue, respectively.}
  \label{fig:diff_flux}
\end{figure}


In Fig.~\ref{fig:diff_flux}, we show the expected ALP flux reaching Earth from galactic SNe as a function of the ALP energy. In the left plot, we fix the ALP mass to $m_a=20$~MeV and took three different values for the coupling in the trapping regime: $g_a=5\times 10^{-6}, \, 10^{-5}, \, 5\times 10^{-5}$, in dotted, dashed-dotted, and solid lines, respectively. Note that $g_{ap}=g_a \, C_{ap} $. In the right plot, we fixed the ALP-proton coupling to $g_a=10^{-5}$, and considered three different values for the ALP mass: $m_a=2, \, 20, \, 65$ MeV, in orange, green and blue, respectively.

To understand the behaviour of the ALP flux shown in Fig.~\ref{fig:diff_flux}, it is helpful to compare it to the well-studied case of massless ALPs in the free-streaming regime. In that case, the ALP production spectrum is expected to peak at  $E_a \sim 50$ MeV (local ALP energy in the proto-NS)~\cite{Carenza:2019pxu, Lella:2023bfb, Carenza:2023lci}. If $g_a$ is large enough for absorption effects to become important, ALPs are diffusively trapped within the proto-NS, and the expected peak moves towards lower energy values, see Refs.~\cite{Lella:2023bfb, Carenza:2023lci} for more details. For $m_a > 1$~MeV, the peak of the spectrum becomes smaller and is located at higher energies compared to the massless case (it moves towards higher energy values for higher ALP masses). This peak can be seen in Fig.~\ref{fig:diff_flux}, after redshift effects have been included in order to get the flux of ALPs as a function of the ALP energy arriving at Earth, i.e., $E_a^{\rm Earth}$. Besides, for $g_a \gtrsim 10^{-7}$, the pionic-ALP conversion peak of the ALP spectrum, expected in the free-streaming regime at $E_a \sim 200$~MeV, washes out because of pionic re-absorption. This is the reason why only one peak is observed in Fig.~\ref{fig:diff_flux}.

From the left plot of Fig.~\ref{fig:diff_flux}, we observe that, in the trapping regime, the ALP flux decreases when the ALP-proton coupling increases. This is because absorption effects become stronger as the coupling constant increases. On the right plot, we can observe the effect of the ALP mass on the ALP flux received at Earth. Specifically, the flux decreases when the mass increases. This is a combined effect of the ALP production spectrum mass dependence, which decreases for higher masses, and the gravitational redshift effect, which is particularly important when $E_a\simeq m_a$, since the kinetic energy of the ALP at the moment of production may not be sufficient enough to scape the star's gravitational well.

\section{Signal in neutrino water Cherenkov detectors}
\label{sec:cherenkov}

\begin{figure}[!t]
  \centering
  \includegraphics[width=.38\textwidth]{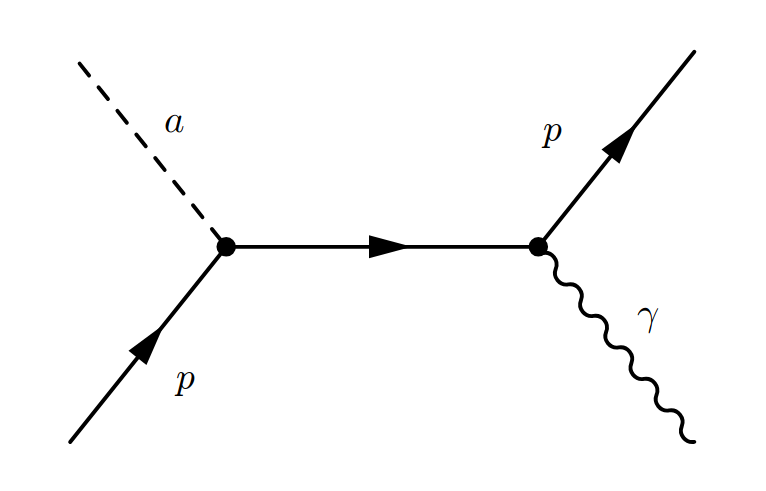}
  \includegraphics[width=.38\textwidth]{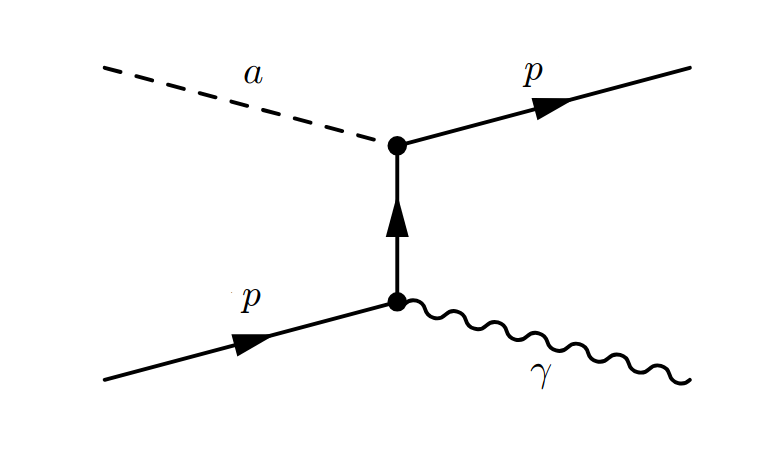}
  \caption{ Feynman diagrams of the detection process.}
  \label{fig:feynman}
\end{figure}

Once we have estimated the flux of ALPs from galactic SNe that reaches Earth, we can study its possible detection in large underground water Cherenkov neutrino experiments, such as SK or the proposed HK. If MeV ALPs couple to protons, the ALP spectrum from galactic SNe can be detected through its interaction with free protons yielding the emission of a photon, as shown in the diagrams of Fig.~\ref{fig:feynman}.

The number of photons per energy and time produced at an experiment with $N_t$ targets can be computed as
\begin{equation}
\frac{d\, N_\gamma}{dE_\gamma}=N_t \int_{E_a^{\rm min}(E_\gamma)}^{E_a^{\rm max}(E_\gamma)} dE_a^{\mathrm{Earth}}\frac{d\Phi_a}{dE_a^{\mathrm{Earth}}} \frac{d\, \sigma_{ap}}{d\, E_\gamma},
\label{eq:photon_spectrum}
\end{equation}
where
\begin{equation}
    \frac{d\sigma_{ap}}{dE_\gamma}= \frac{1}{32\pi} \int_{-1}^{1} \frac{\left|\overline{\mathcal{M}}\right|_{ap}^2}{\left|\vec{p_a}\right|^2 m_p}\ \delta(\cos\theta-\cos\theta_0)\ d\textrm{cos}\,\theta
    \label{eq:crosssection}
\end{equation}
is the ALP-proton cross section,
\begin{align}
 |\overline{\mathcal{M}}|_{ap}^2= & \frac{C_{ap}^2 e^2 g_a^2}{E_\gamma^2 m_p \left(2 E_a m_p+m_a^2\right)^2} \; \bigg[ 4 m_p^3 (E_a-E_\gamma)^2 \left(2 E_a E_\gamma+m_a^2\right) \notag \\ & +4 m_a^2 m_p^2
		(E_a-E_\gamma) \left(E_\gamma (E_a-E_\gamma)+m_a^2\right) \notag \\  &+ m_a^4 m_p \left(2 E_a E_\gamma
		+m_a^2\right)+E_\gamma m_a^6 \bigg]
\end{align}
is the averaged squared amplitude of the process, and
\begin{equation}
    \cos\theta_0 = \frac{2E_\gamma(m_p+E_a)-2m_pE_a-m_a^2}{2E_\gamma\sqrt{E_a^2-m_a^2}}
\end{equation}
is the cosine of the angle between the photon and the ALP momenta, which is fixed by energy conservation. Imposing $|\cos \theta_0| \leq 1$, sets the minimum and maximum allowed energies for the photon produced by an ALP with energy $E_a$,
\begin{equation}
   E_\gamma^{\rm min}=\frac{m_a^2 + 2 E_a m_p}{2 (m_p + E_a) + 2 \sqrt{E_a^2 - m_a^2}},
	\,\,\,\,\,\,\,\,E_\gamma^{\rm max}=\frac{m_a^2 + 2 E_a m_p}{2 (m_p + E_a) - 2 \sqrt{E_a^2 - m_a^2}}. 
    \label{eq:elimits}
\end{equation}


In our scenario, $1\,\textrm{MeV}<E_a<100\,\textrm{MeV}$, therefore, $10^{-3} \, \textrm{MeV} <E_\gamma^{\rm max}-E_\gamma^{\rm min} < 18$~MeV (upper limit for the highest ALP energy). Similarly, $E_a^{\rm min}$ and $E_a^{\rm max}$ as a function of the photon energy can be obtained, allowing us to perform the integral in Eq.~\eqref{eq:photon_spectrum}.

\subsection{Constraints from Super-Kamiokande}

\begin{figure}[!t]
  \centering
  \includegraphics[width=.55\textwidth]{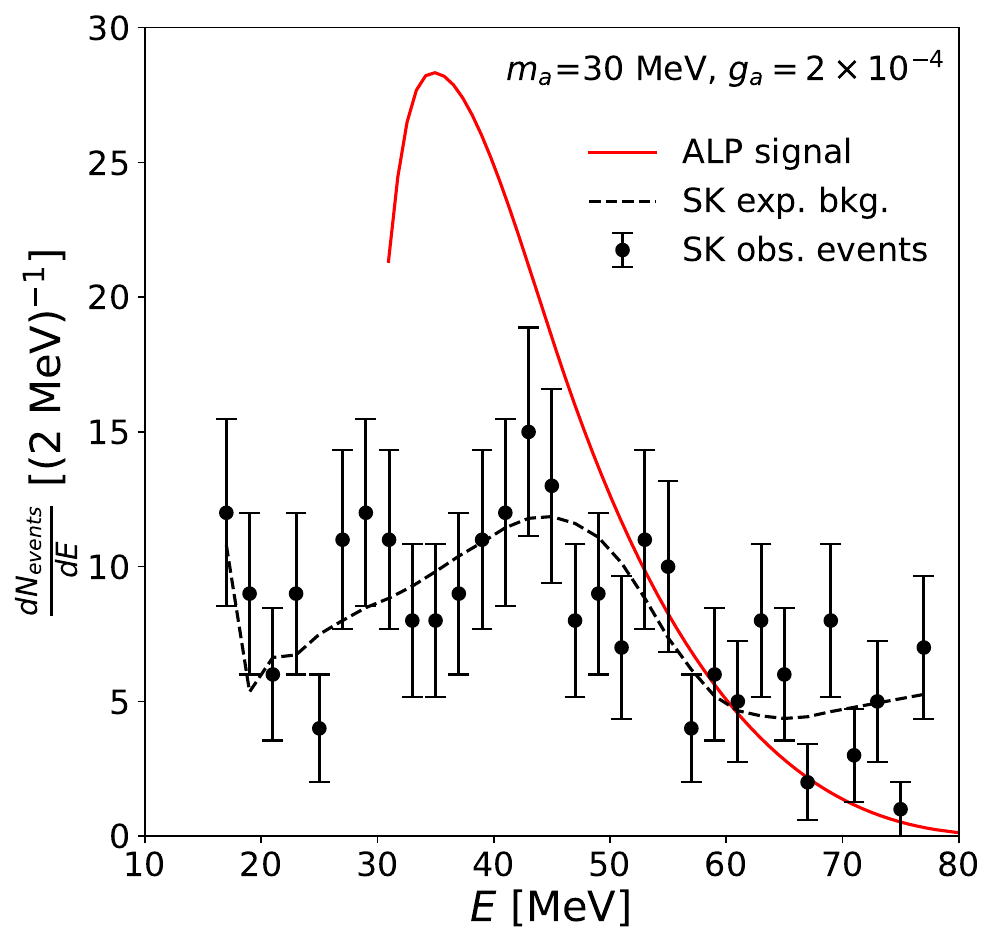}

  \caption{Expected event rate in Super-Kamiokande phase IV via $a \, p \rightarrow p \, \gamma$ (red solid line) for $g_a=2 \times 10^{-4}$ and $m_a=30$ MeV and for the expected events from SM processes taken from~\cite{Super-Kamiokande:2021jaq} (black dashed line). The observed events are shown with their error bars. 
  }
  \label{fig:events}
\end{figure}

Fig.~\ref{fig:events} shows the expected event rate at SK phase IV (exposure $22.5 \times 2970$~kton~days) from $a \, p \rightarrow p\, \gamma$ interactions for $g_a=2 \times 10^{-4}$ and $m_a=30$~MeV with a red solid line. The expected background and observed events (in black) correspond to those in the signal region without neutron coincidence (see Fig. 26 of Ref.~\cite{Super-Kamiokande:2021jaq}). The ALP signal peaks at photon energies $\sim 35$ MeV, for which Eq.~\eqref{eq:elimits} predicts $E_\gamma \sim E_a$. The signal for this example clearly exceeds SK observation and would therefore be excluded.

The reconstructed energy region in Fig.~\ref{fig:events}, $E_{rec}=[16, 80]$ MeV, was optimized to search for the DSNB in Ref.~\cite{Super-Kamiokande:2021jaq} by identifying positrons from inverse beta decay interactions. Since they are ultra-relativistic, the Cherenkov signal produced by one positron is indistinguishable from that generated by a photon. We apply the same signal region to search for photons produced by ALP interactions. The main background in this region is dominated by cosmic ray muon spallation, electrons produced by the decays of invisible, or low energy, muons and pions, and atmospheric neutrinos for the lower, middle and higher sections of the signal energy range, respectively~\cite{Super-Kamiokande:2021jaq}.

For higher energies, the flux of ALPs in the trapping regime decreases significantly and we expect no signal above $\sim 100$ MeV. For lower energies we do expect a non-negligible signal for $m_a\lesssim 15$ MeV, however for $E_{rec}< 15$ MeV the expected background increases by a few orders of magnitude, dominated by spallation and accidental coincidences events. Additionally, the event rate spectrum for $m_a \lesssim 30$ MeV peaks at $E_{rec} \simeq 20-30$ MeV, were the signal efficiency is close to 100$\%$ (see Fig. 18 of Ref.~\cite{Super-Kamiokande:2021jaq}) making the $E_{rec}=[16, 80]$ MeV energy region ideal to probe our scenario.

\begin{figure}[!t]
  \centering
  
  \includegraphics[width=.55\textwidth]{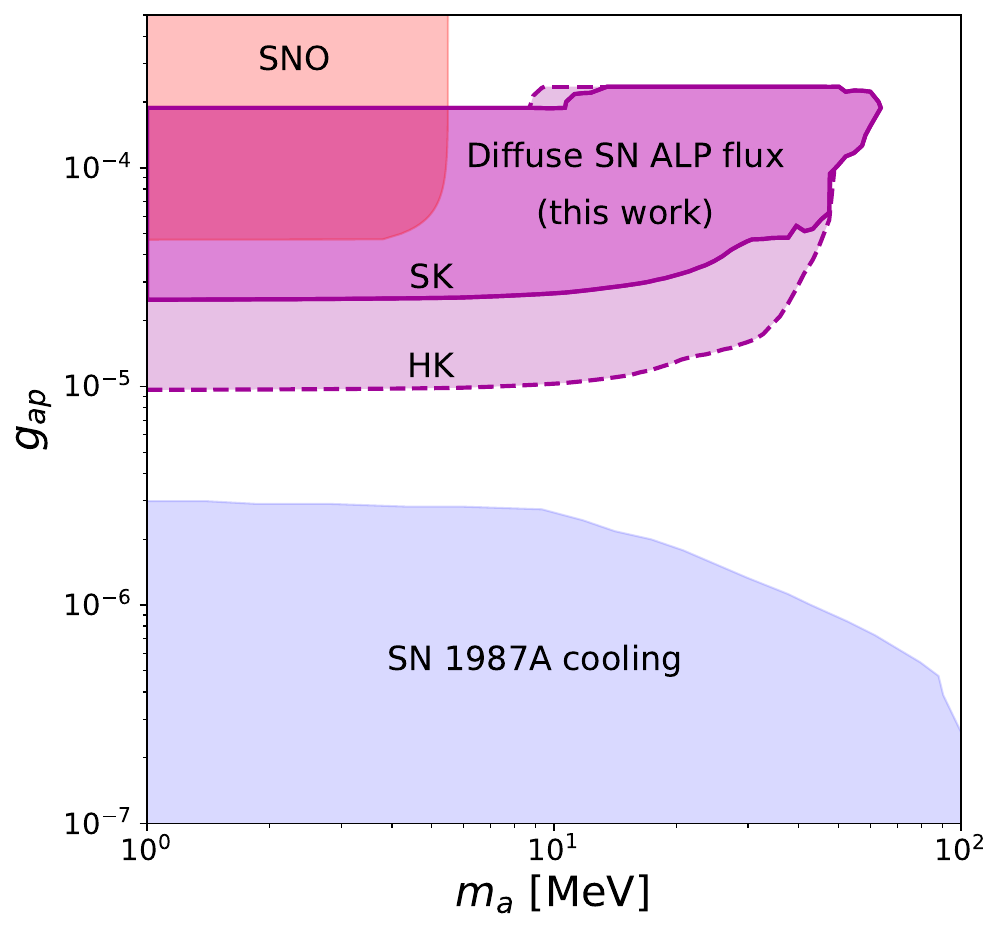}
  \caption{Bounds on the ALP parameter space, ALP-proton coupling constant vs ALP mass. The pink region is excluded by the expected events in SK of ALPs from the galactic diffuse SN flux scattering on free protons. The light pink region is the projected excluded region for HK. Complementary bounds from expected solar axion events in SNO \cite{Bhusal:2020bvx} (red region) and from SN 1987A cooling \cite{Lella:2023bfb} (blue region). 
  }
  \label{fig:bounds}
\end{figure}

SK data is consistent with the expected background, and therefore we can use it to set constraints on the ALP signal.
Fig.~\ref{fig:bounds} shows (in pink) the resulting excluded region in the $(g_{ap},\,m_a)$ parameter space from the data of the first four SK phases (exposure $22.5 \times 5823$~kton~days)~\cite{Super-Kamiokande:2021jaq}. To derive this region, we have compared the observed number of events with the expected signal and background yields through a profiled log-likelihood ratio test, a binned analysis in the reconstructed energy signal range mentioned before, with bin width of 2 MeV (notice that each SK phase has different exposure time and slightly different $E_{rec}^{\rm min}$ and $E_{rec}^{\rm max}$). Although the bound shown was obtained combining the data from the four SK phases, the result is dominated by SK-IV information, given its larger exposure. For further details about the statistical treatment, see Appendix~\ref{app:stats}.

In lighter pink, we show the region that will be probed by HK with an exposure of $187\times 10$~kton~years. We have taken the expected background from the HK report, see Ref.~\cite{Hyper-Kamiokande:2018ofw}, where $E_{rec}=[16, 50]$ MeV, therefore the projected exclusion can only probe ALPs with $m_a \leq 50$ MeV. For comparison, in red and blue complementary bounds from expected solar axion events in SNO \cite{Bhusal:2020bvx} and from SN 1987A cooling \cite{Lella:2023bfb}, respectively, are shown.

As shown in Fig.~\ref{fig:bounds}, our analysis excludes a region above the cooling bounds, covering one order of magnitude in the ALP-proton coupling, between $\sim 2 \times 10^{-5}$ and $\sim 2 \times 10^{-4}$, for $m_a \sim 1-70$ MeV. The projected sensitivity of HK strengthens the constraints from SK by up to a factor of $\sim 2.5$ for $m_a \sim 1-50$ MeV.
For couplings above $g_{ap} \sim 2 \times 10^{-4}$, absorption effects become so significant that ALPs are unable to escape the proto-NS core, preventing their arrival at Earth.
The excluded area is limited to $m_a\lesssim 70$~MeV. The reason is that our limits are dominated by the analysed SK-IV dataset, which comprises events with energies up to $80$~MeV. Additionally, as the ALP mass increases, the production spectrum becomes increasingly suppressed, entirely vanishing for $m_a\gtrsim 200$ MeV, since the SN environment lacks the energy required to produce such heavy particles. 
Finally, we do not consider ALP masses below $1$ MeV, since the assumption of a diffuse ALP flux begins to break down. 



Finally, note that the process $ a \, p \rightarrow p \, \gamma $ under consideration is not the only possible search in neutrino Cherenkov detectors. Previous studies \cite{Engel:1990zd, Carenza:2023wsm} have suggested looking for ALP-induced excitations of oxygen nuclei by detecting the photons emitted during their de-excitation. In such a case, the observed experimental signal would consist primarily of photons distributed in the energy range $[5,10]$~ MeV, corresponding to transitions between the most likely excited states of oxygen and its ground state. This search could also be used to determine exclusion limits in the region of the parameter space we are considering. By performing a simple extrapolation and rescaling of the results obtained in Ref.~\cite{Carenza:2023wsm} (taking into account that their analysis assumes ALPs coming from SN 1987A) we estimate that the number of photons emitted due to oxygen de-excitation would exceed those from our detection channel by two orders of magnitude. However, in the energy range where the de-excitation photons are expected, the background events also increase by a few orders of magnitude, dominated by spallation and accidental coincidences, as mentioned before. Thus, a dedicated background reduction study, taking into account the detector characteristics near its energy threshold, and a proper signal prediction for massive ALPs, would be required to assess the viability of the ALP-oxygen signal. Although interesting, this analysis is beyond the scope of this work.

\section{Conclusions}\label{sec:conclusions}

In this article, we claim that axion-like particles (ALPs) with masses in the MeV range coupled to nucleons can be produced with semi-relativistic velocities in core-collapse supernovae (SNe), which would give rise to a diffuse galactic flux. We show that these ALPs can be detected in neutrino water Cherenkov detectors and produce photons with energies of $\mathcal{O}(10)$ MeV, where the experimental background is very small. This allows us to put stringent constraints on the ALP-proton coupling for masses in the range $1-70$~MeV.

In particular, we have computed the spectral fluence of ALPs reaching Earth from core-collapse SNe. This calculation incorporates key effects such as ALP reabsorption through $N\, N\, a \rightarrow N\, N$ and $N\, a \rightarrow N\, \pi$, which reduce the ALP flux in certain energy ranges, as well as the gravitational redshift caused by the proto-NS gravitational potential, both within and outside the SN environment. Using the spectral fluence, we derived the diffuse galactic ALP flux arriving at Earth by weighting the fluence with the galactic SN rate. This approach assumes a near-constant galactic flux, which is valid for $m_a \gtrsim 1$ MeV, where ALPs reach Earth in a time window $\gtrsim 5 \times 10^2$ years.

We then calculated the number of photon events as a function of energy at SK and HK for a benchmark model where the ALP-neutron coupling is negligible. Comparing these signals with observed and expected background, we found that for $m_a = 1-70$~MeV and ALP-proton couplings in the range $g_{ap} \sim 2 \times 10^{-5} - 2 \times 10^{-4}$, the expected signal at SK exceeds the observed background, allowing us to place robust bounds on the ALP parameter space. Projected constraints from HK improve the limits of SK by a factor of $\sim 2.5$.

The constraints set in this work, which span one order of magnitude above cooling bounds, probe the trapping regime of MeV ALPs for the first time by considering the diffuse galactic ALP flux from SNe. 

\vspace{2.5mm}
\section*{Acknowledgments.}

We would like to thank Pierluca Carenza, William DeRocco, Luis Labarga, Gabriel Martínez-Pinedo, Daniel Naredo Tuero, Mario Reig and Javi Serra for useful discussions and comments. We also thank Nick Houston for details on the limits from SNO. We acknowledge support from the Comunidad Autónoma de Madrid and Universidad Autónoma de Madrid under grant SI2/PBG/2020-00005, and by the Spanish Agencia Estatal de Investigaci\'on through the grants PID2021-125331NB-I00 and CEX2020-001007-S, funded by MCIN/AEI/10.13039/501100011033. DGC also acknowledges support from the Spanish Ministerio de Ciencia e Innovaci\'on under grant CNS2022-135702.

\section*{Appendix}
\appendix

\section{ALP photo-production in the SN interior}
\label{app:production}

In order to show that the ALP production rate in the SN interior through the process $\gamma \, p\rightarrow p\, a$ is negligible compared to the one via nucleon-nucleon Bremmstrahlung and pionic conversion, in this appendix, we estimate the ALP production spectrum via $\gamma \, p\rightarrow p\, a$ inside the proto-NS. Using the general formula of Eq.\eqref{eq:prod_spectrum}, we can write the ALP production spectrum for ALP photo-production as 
\begin{align}
    \frac{d^2n_a}{dE_a dt} = & \int g_\gamma\frac{d^3\vec{p}_\gamma f_{\gamma}(E_\gamma)}{(2\pi)^32E_\gamma}\int g_p\frac{d^3\vec{p}_{p_1} f_p(E_{p_1})}{(2\pi)^32E_{p_1}}\int \frac{d^3\vec{p}_{p_2}}{(2\pi)^32E_{p_2}}(1-f_p(E_{p_2})) \nonumber \\
    & \int \frac{d\cos\theta_a}{2} \frac{|{\vec{p}_a}|}{4 \pi^2} |{\mathcal{\overline{M}}|}^2(2\pi)^4\delta^{(4)}(p_{p_1}+p_\gamma-p_{p_2}-p_a), 
\end{align}
where $g_\gamma=3$ and $g_p=2$ are the photon and proton degrees of freedom, respectively,  $f_\gamma(E_\gamma)$ is the Bose-Einstein distribution function for a photon with energy $E_\gamma$ and effective mass $m_\gamma$, and $f_p(E_p)$ is the Fermi-Dirac distribution function for protons with effective mass $m_p^*$ and effective chemical potential $\mu_p^*$, see Ref.~\cite{Carenza:2019pxu, Lucente:2020whw}. Since $m_p^* \sim 0.5-1$ GeV (lowest value for the inner-most part of the proto-NS), $m_p^* \gg m_\gamma \sim 18$~MeV~\cite{Ferreira:2022xlw} and $m_p^* \gg T\sim 30$~MeV, we can assume $E_{p_1} \simeq E_{p_2}$. Therefore, $f_p(E_{p_2}) \simeq f_p(E_{p_1}) $ and we can factorise the following quantity
\begin{align}
    \frac{1}{2E_\gamma2E_{p_1}}\int \frac{d^3\vec{p}_{p_2}}{(2\pi)^32E_{p_2}}& \int \frac{d\cos\theta_a}{2} \frac{|{\vec{p}_a}|}{4 \pi^2} |{\mathcal{\overline{M}}|}^2(2\pi)^4\delta^{(4)}(p_{p_1}+p_\gamma-p_{p_2}-p_a) \nonumber \\
    & = \frac{\sqrt{(p_\gamma p_{p_1})^2-m_\gamma^2m_p^{*2}}}{E_\gamma E_{p_1}} \, \frac{d\sigma}{dE_a} \nonumber \\
    & \simeq \frac{\sqrt{E_\gamma^2-m_\gamma^2}}{E_\gamma} \, \frac{d\sigma}{dE_a} ,
\end{align}
where in the last step we used $E_{p_1} \simeq m_p^*$.
The differential cross section for this process can be obtained as
\begin{equation}
    \frac{d\sigma}{dE_a} = \frac{1}{32\pi} \int_{-1}^{1} \frac{\left|\overline{\mathcal{M}}\right|^2}{\left|\vec{p_\gamma}\right|^2 m_p^*}\ \delta(\cos\theta_a-\cos\theta^0_a)\ d\cos\theta_a,
\end{equation}
with 
\begin{equation}
    \cos\theta^0_a=\frac{E_\gamma E_a-m_p^*(E_\gamma-E_a)-\frac{1}{2}(m_a^2+m_\gamma^2)}{\sqrt{E_\gamma^2-m_\gamma^2}{\sqrt{E_a^2-m_a^2}}}
\end{equation}
determined by energy conservation, and with the averaged squared amplitude given by
\begin{align}
    \left|\overline{\mathcal{M}}\right|^2 = & \frac43C_{ap}^2e^2g_a^2m_p^*\left[\frac{2E_a(m_a^2+E_\gamma m_p^*)+m_a^2(m_p^*-E_\gamma)-4E_a^2m_p^*+E_am_\gamma^2}{(m_a^2-2E_am_p^*)^2}\right. \nonumber \\
    & \left.+\frac{m_a^2(m_p^*-E_\gamma)}{(2E_\gamma m_p^*+m_\gamma^2)^2}+\frac{2m_a^2m_p^*-2E_a^2m_p^*-E_am_a^2}{(m_a^2-2E_am_p^*)(2E_\gamma m_p^*+m_\gamma^2)}\right].
\end{align}

Then, the spectrum reads 
\begin{equation}
    \frac{d^2n_a}{dE_a dt} \approx \int g_\gamma\frac{d^3\vec{p}_\gamma f_\gamma(E_\gamma)}{(2\pi)^3}\int g_p\frac{d^3\vec{p}_{p_1} f_p(E_{p_1})}{(2\pi)^3}(1-f_p(E_{p_1}))\frac{\sqrt{E_\gamma^2-m_\gamma^2}}{E_\gamma}\frac{d\sigma}{dE_a}.
\end{equation}

As it has been done in Ref.\cite{Ferreira:2022xlw}, we define the effective number density of protons as
\begin{equation}
    n_p^\text{eff}\equiv 2 \int \frac{d^3\vec{p}_{p_1} f_p(E_{p_1})}{(2\pi)^3}(1-f_p(E_{p_1})),
\end{equation}
so that we can write
\begin{align}
    \frac{d^2n_a}{dE_a dt} & \approx n_p^\text{eff}\int 3\frac{d^3\vec{p}_\gamma}{(2\pi)^3}\frac{1}{e^{E_\gamma/T}-1}\frac{\sqrt{E_\gamma^2-m_\gamma^2}}{E_\gamma}\frac{d\sigma}{dE_a} \nonumber \\ & =n_p^\text{eff}\frac{6}{(2\pi)^2}\int dE_\gamma\frac{1}{e^{E_\gamma/T}-1}(E_\gamma^2-m_\gamma^2)\frac{d\sigma}{dE_a}.
    \label{eq:photoprod}
\end{align}

\begin{figure}[!t]
  \centering
  \includegraphics[width=.505\textwidth]{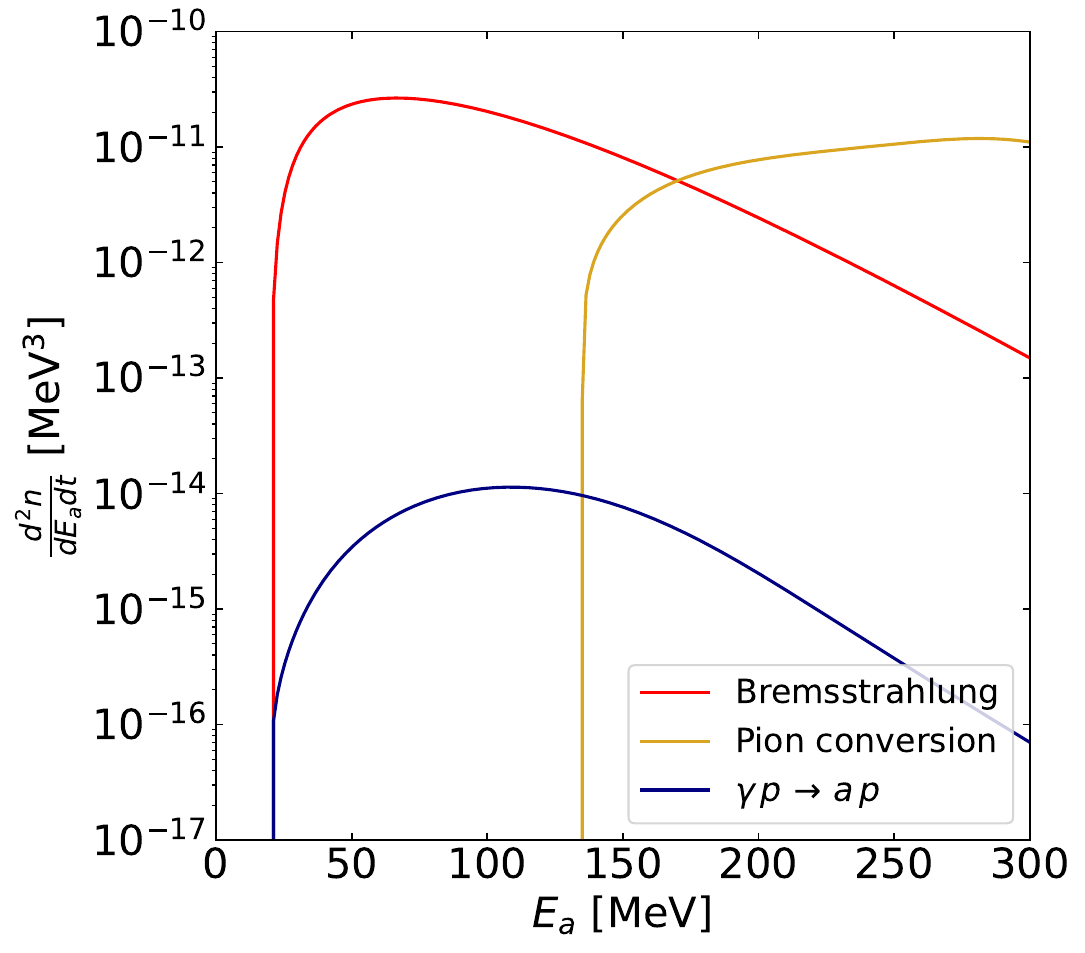}
  \includegraphics[width=.485\textwidth]{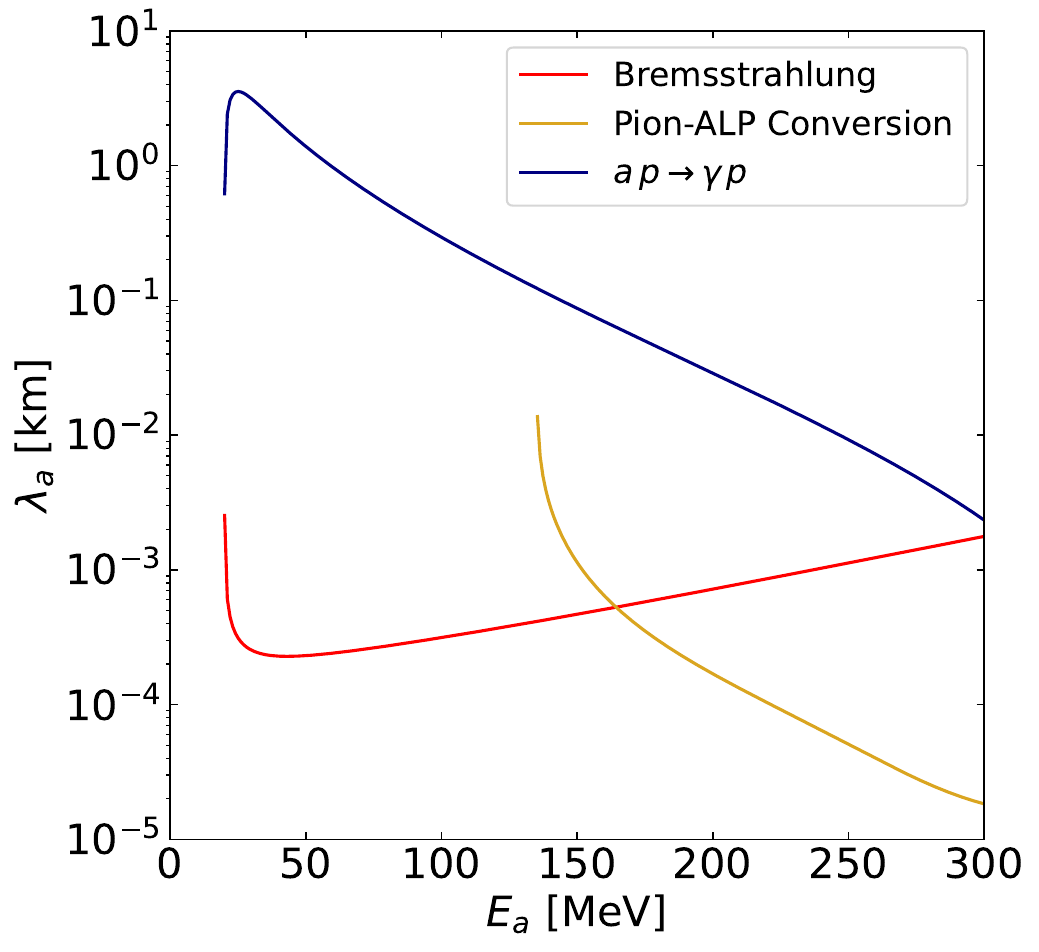}
  \caption{ALP production spectrum (left panel) and mean free path (right panel) from nucleon-nucleon bremsstrahlung (red), pion conversion (mustard) and $\gamma\, p\rightarrow a\, p$ (blue) processes at radius $R=10$~km for an ALP with mass $m_a=20$~MeV and coupling $g_a=10^{-5}$.}
  \label{fig:contributions}
\end{figure}

On the left panel of Fig.~\ref{fig:contributions} we show the ALP production spectrum inside the proto-NS at a fixed radius $R=10$~km for an ALP with mass $m_a=20$~MeV and coupling $g_a=10^{-5}$ (fixing $C_{ap}=-0.47$ and $C_{an}=0$) from nucleon-nucleon bremsstrahlung (red line), pion conversion (yellow) and ALP photo-production (green). For bremsstrahlung and pion conversion we have used the expressions given in Ref.~\cite{Carenza:2023lci}, while for ALP photo-production we have used the result obtained from Eq.~\eqref{eq:photoprod}. We have considered the temperature, density, electron and muon fraction profiles of Ref.~\cite{Fischer:2021jfm} at 1 second after bounce, and the values of the effective photon mass and the effective density of protons of Ref.\cite{Ferreira:2022xlw} (which were obtained using the profiles of Ref.~\cite{Fischer:2021jfm}). For illustration, we fix the radius at $R=10$~km, where the production is expected to be maximal for the used profiles. 

It is clear that the ALP production spectrum due to ALP photo-production is negligible compared to the one due to bremsstrahlung and pion conversion. Note that the three contributions to the spectrum depend quadratically on the coupling $g_a$, therefore this result can be easily extrapolated to any value of the coupling constant. 

Finally, on the right panel of Fig.~\ref{fig:contributions}, we show the mean free path for the same processes and the same parameters as discussed above. Since the mean free path for $a\, p\rightarrow \gamma \, p$ is orders of magnitude larger than bremsstrahlung and pion conversion processes, it does not affect the computation for the absorption effect in Eq.~\eqref{eq:spectrum}.

\section{Statistical analysis}
\label{app:stats}
We present in this appendix the details of the statistical analysis performed in this work to obtain the observed and expected exclusion limits. We considered the profiled log-likelihood-ratio test statistic, defined as
\begin{equation}
    q(\theta) = -2 \, \ln \left( \frac{\mathcal{L}(\theta)}{\mathcal{L}(\hat{\theta})} \right) ,
\end{equation}
where $\mathcal{L}$ is the likelihood function that describes the data given the model parameters $\theta$, and $\hat{\theta}$ indicate the quantities that maximise the likelihood.

For Super-Kamiokande we performed a binned statistical treatment assuming that the number of events in each bin follows a Poisson distribution due to the lower number of events expected in some bins. Thus, the test statistics becomes
\begin{equation}
    q(\theta) = 2 \left[ \sum_{i=1}^{n_{bins}} \, N^i_{th}(\theta) \, - \, N^i_{\rm obs} \, + \, N^i_{\rm obs} \, \ln \left( \frac{N^i_{\rm obs}}{N^i_{th}(\theta)} \right) \right],
\end{equation}
with $N^i_{\rm obs}$ and $N^i_{th}(\theta)$ the number of observed and expected events in the bin $i$, and $n_{bins}$ the total number of bins. For exclusion, $N^k_{th}(\theta)$ includes both signal and background contributions. The data was taken from Ref.~\cite{Super-Kamiokande:2021jaq}. To compute the number of expected signal events we considered $22.5$~kton active detector material and the data from the first fourth operational phases: SK-I and SK-III with $n_{bins}=36$, bin width of $2$ MeV in a range $[16,88]$~MeV and $1497$ and $562$ days of exposure, respectively, SK-II with $n_{bins}=36$, bin width of $2$ MeV in a range $[17.5, 89.5]$~MeV and $794$ days of exposure, while for SK-IV, $n_{bins}=31$, bin width of $2$ MeV in a range $[16, 78]$~MeV and $2970$~days of exposure. The background takes into account the Horiuchi+09~\cite{Horiuchi:2008jz} DSNB.

\begin{figure}[!t]
  \centering
  
  \includegraphics[width=.55\textwidth]{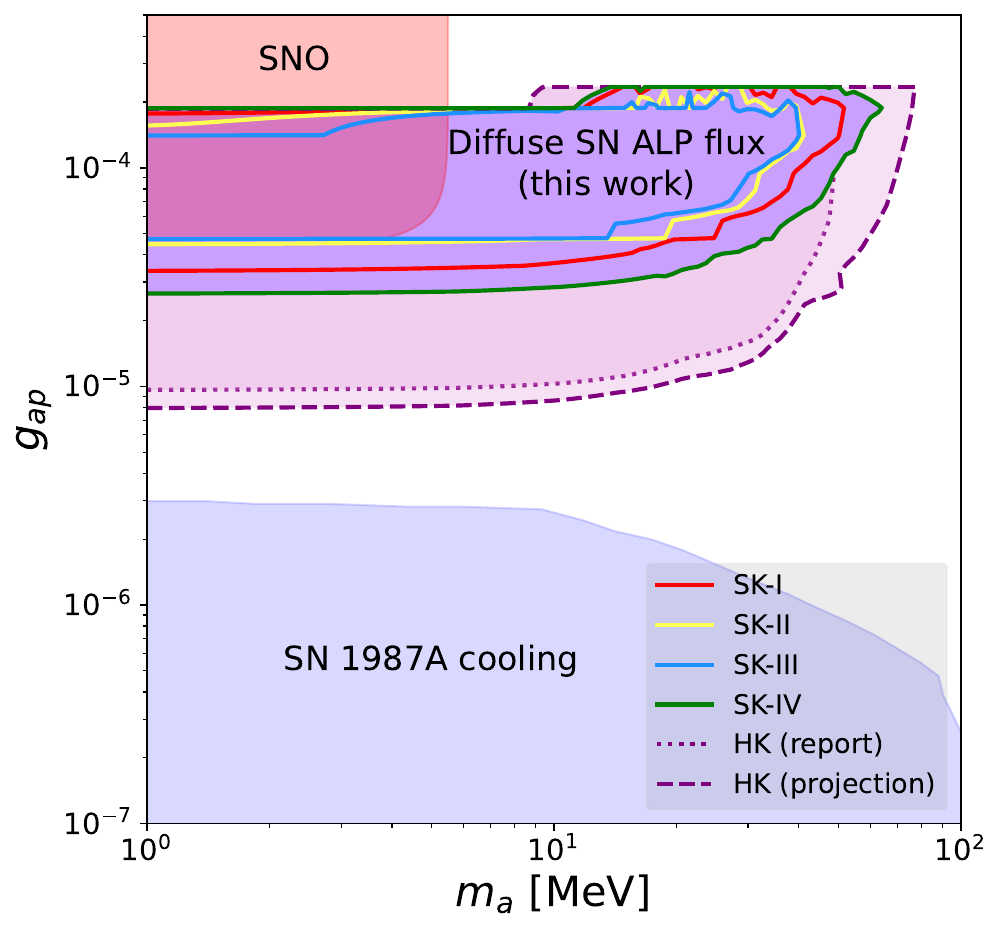}
  \caption{Same as Fig.~\ref{fig:bounds}, but for each SK phase independently and HK using the expected background published in a design report~\cite{Hyper-Kamiokande:2018ofw} or the projected background assuming the same response and data reduction used in a recent SK analysis~\cite{Super-Kamiokande:2021jaq}.
  }
  \label{fig:bounds_compare}
\end{figure}

To compute the 95\% C.L. we used the Wilks’ theorem, which establishes that the log-likelihood-ratio test statistic $q(\theta)$ asymptotically follows a $\chi^2$ distribution with number of degrees of freedom equal to the difference between the number of bins and the number of free parameters of the model, the ALP mass and coupling. For SK-I, SK-II and SK-III this means $q(\theta)=48.6$, while for SK-IV $q(\theta)=42.6$. We show the bounds in Fig.~\ref{fig:bounds_compare}. As expected by the longer exposure time, we found that the constraint set by SK-IV data is the strongest one.

In order to combine the four phases of SK, we compared two methods. First, we assumed that each one is an independent experiment, then the total $q^{\rm comb}(\theta) = \sum_i^4 q^{i}(\theta)$ described by a $\chi^2$ distribution with number of degrees of freedom $dof^{\rm comb}= \sum_i^4 dof^{i}$, with the index $i=1,..4$ representing each SK phase. As a second approach, we built a single experiment combining the exposure of the four SK phases, i.e. combining the number of observed events and expected background, and computed a new $q^{\rm comb}(\theta)$ the using the same binning as in SK-IV and therefore the same number of degrees of freedom. The exclusion limits found with both methods present no significant differences, and are only slightly more restrictive than the limits set by only SK-IV data.

To determine the projected exclusion limits for HK we considered
\begin{equation}
    Z(\theta) = \left[2 \sum_{i=1}^{n_{bins}} \, S^i_{th}(\theta) \, + \, B^i_{th} \, \ln \left( \frac{B^i_{th}}{S^i_{th}(\theta) \, + \, B^i_{th}} \right) \right]^{1/2},
\end{equation}
using Asimov data sets, meaning that the observed data is set to the theoretically expected number of events. Here $S^i_{th}(\theta)$ and $B^i_{th}$ are the expected number of signal and background events in each bin $i$. We stablished the 95\% C.L. at $Z=1.96$. The information about the expected background, which includes the DSNB model, was taken from the Hyper-Kamiokande report~\cite{Hyper-Kamiokande:2018ofw}, with an exposure of $187\times 10$~kton years, $n_{bins}=17$, and bin width of $2$ MeV in a range $[16, 50]$~MeV. If we assume that the characteristics of HK will be the same as SK and that we can apply the same data reduction techniques applied in the more recent SK analysis~\cite{Super-Kamiokande:2021jaq}, we can project the expected backgrounds by rescaling it with the exposure time and detector mass ($n_{bins}=31$, bin width of $2$ MeV in a range $[16, 78]$~MeV). In Fig.~\ref{fig:bounds_compare} we show the comparison of both approaches, and we can see that although the former leads to a more conservative constraint, both can probe the parameter region between current cooling and diffuse SN ALP bounds.


\bibliographystyle{JHEP} 
\bibliography{diffuse_main_correction.bbl}

\end{document}